\documentclass[aps,prl,reprint,superscriptaddress,longbibliography]{revtex4-2}

\usepackage[english]{babel}
\usepackage{graphicx}
\usepackage{amsmath,amssymb}
\usepackage{comment}
\usepackage{dcolumn}
\usepackage[T1]{fontenc}

\usepackage[
    colorlinks=true,
    linkcolor=blue,
    citecolor=blue,
    urlcolor=blue
]{hyperref}

\usepackage{todonotes}

\begin{document}

%\preprint{}

\title{Evidence for an Obstruction to Perturbative Renormalization Group in Stochastic Turbulence}


\author{L. Ts. Adzhemyan}
\affiliation{Saint Petersburg State University, 7/9 Universitetskaya nab., St. Petersburg, 199034, Russian Federation}
\affiliation{BLTP, Joint Institute for Nuclear Research, 
 Dubna, Moscow Region, 141980,  Russian Federation}

\author{D. A. Davletbaeva}
\affiliation{Saint Petersburg State University, 7/9 Universitetskaya nab., St. Petersburg, 199034, Russian Federation}
\affiliation{Petersburg Nuclear Physics Institute named by B. P. Konstantinov of NRC ``Kurchatov Institute'', mkr. Orlova roshcha 1, Gatchina, Leningradskaya oblast, 188300, Russian Federation}

\author{D. A. Evdokimov}
\affiliation{Saint Petersburg State University, 7/9 Universitetskaya nab., St. Petersburg, 199034, Russian Federation}
\affiliation{BLTP, Joint Institute for Nuclear Research, 
 Dubna, Moscow Region, 141980,  Russian Federation}

\author{M. V. Kompaniets}
\email{m.kompaniets@spbu.ru}
\affiliation{Saint Petersburg State University, 7/9 Universitetskaya nab., St. Petersburg, 199034, Russian Federation}
\affiliation{BLTP, Joint Institute for Nuclear Research, 
 Dubna, Moscow Region, 141980,  Russian Federation}


\date{\today}

\begin{abstract}
We examine whether the perturbative renormalization group (RG) regime of stochastic turbulence can be continued from small $\varepsilon$ to the physical value $\varepsilon_{\mathrm{phys}}=2$, corresponding to large-scale forcing. Exploiting the simplifications of the large-$d$ limit, we calculate the correction exponent $\omega$ to five-loop order and find a simple structure suggesting a closed analytic form. This form implies that the fixed point stable near $\varepsilon=0$ ceases to be simultaneously real and infrared stable before reaching $\varepsilon_{\mathrm{phys}}=2$, pointing to a possible change of infrared regime rather than a controlled perturbative RG extrapolation.
\end{abstract}

%\keywords{}

\maketitle

%\subsection{}

\textit{Introduction.---} The statistical properties of fully developed turbulence and a rigorous explanation of deviations from Kolmogorov’s phenomenological theory remain among the central unresolved problems of classical physics~\cite{K41,Frisch1995,Falkovich2001}. Substantial progress in this direction has been associated with the application of the renormalization group (RG), which provided a foundation for several phenomenological concepts and, within the perturbative $\varepsilon$-expansion, a systematic method for calculating universal quantities~\cite{Forster1977,Fournier1983,Adzhemyan1996,Adzhemyan1999}. At the same time, it remains an open question whether the perturbative RG (PRG) can be continuously extended from the controlled vicinity of $\varepsilon=0$ to the physical value $\varepsilon_{\mathrm{phys}}=2$~\cite{Adzhemyan2003,Zhou2010,FRG_Canet2022}. In this Letter, we investigate this question in the limit of large spatial dimension $d$ and show that the five-loop calculation of the correction-to-scaling exponent $\omega$ indicates a possible obstruction to such a continuation.

The standard field-theoretic formulation of homogeneous isotropic stochastic turbulence~\cite{Forster1977,Adzhemyan1999,Vasiliev2004} is based on the $d$-dimensional Navier--Stokes equation driven by a Gaussian random force whose scalar spectral function has the power-law form
\begin{equation}
d_{\mathrm f}(k) = k^{4-d-2\varepsilon}.
\label{eq:pump}
\end{equation}
Unlike in conventional RG applications to critical phenomena, the parameter $\varepsilon$ is not related to the spatial dimension but controls the spectral distribution of energy input. The scaling behavior in the inertial range is governed by a nontrivial RG fixed point. The perturbative expansion is constructed near $\varepsilon=0$, where forcing is ultraviolet (UV), whereas physical infrared (IR) forcing concentrated at large spatial scales corresponds to $\varepsilon_{\mathrm{phys}}=2$. The applicability of this approach to real turbulence therefore requires that the fixed point found at small $\varepsilon$ can be continued to $\varepsilon_{\mathrm{phys}}$ without a change of scaling regime. Such a continuation, however, is not guaranteed: large-scale forcing may be governed by a different fixed point from the one perturbatively accessible near $\varepsilon=0$, where the forcing is UV.

A practical route to studying this problem at high orders in perturbation theory is provided by considering the limit $d\to\infty$, in which the diagrammatic structure of the stochastic Navier--Stokes model simplifies substantially~\cite{Fournier1978, Adzhemyan2008}. Large spatial dimension also provides a natural starting point for a double expansion in $\varepsilon$ and $1/d$, since intermittency and anomalous corrections to Kolmogorov scaling are expected to be suppressed as $d\to\infty$~\cite{Kraichnan1974,Frisch1994,Adzhemyan2001,Antonov2024}. Of particular interest is the correction-to-scaling exponent $\omega$: its sign determines the stability of the nontrivial fixed point, while its magnitude governs the leading corrections to scaling. Previous three- and four-loop calculations of $\omega$ to leading order in $1/d$ revealed an unexpectedly simple rational structure of the coefficients in the $\varepsilon$-expansion~\cite{Adzhemyan2008,Adzhemyan2024}. These studies suggested that the series has a finite radius of convergence that may lie below $\varepsilon_{\mathrm{phys}}=2$, possibly near $1/2$. Computing further coefficients provides a way to test this structure and to better identify the range of validity of the PRG.

We analytically compute the five-loop coefficient in the expansion of the exponent $\omega$ in the limit $d\to\infty$:
\begin{equation}
\omega =
2\varepsilon
+\frac{2}{3}\varepsilon^2
+ \frac{10}{9}\varepsilon^3
+\frac{56}{27}\varepsilon^4
+ \frac{340}{81}\varepsilon^5
+\mathcal{O}(\varepsilon^6).
  \label{eq:omega_series}
\end{equation}
The new coefficient preserves the simple arithmetic structure observed at lower orders. Guided by this structure, we consider possible closed-form continuations of the series. The resulting continuation suggests a nontrivial scenario: the fixed point stable at small $\varepsilon$ does not remain real and stable throughout the interval $0<\varepsilon<2$, while a second real stable branch may exist near $\varepsilon_{\mathrm{phys}}$. This points to a possible separation between the perturbative regime with power-law forcing and the physical regime of IR forcing at the integral scale.

This scenario is consistent with the picture emerging from functional RG (FRG) studies, in which UV and IR forcing may lead to different fixed points and universality classes~\cite{FRG_MejaMonasterio2012,FRG_Canet2016,FRG_Canet2022}. The five-loop result therefore represents not only an extension of the large-$d$ RG calculation to higher perturbative order, but also a direct test of the central assumption underlying the application of the $\varepsilon$-expansion to physical turbulence.

\textit{Five-loop RG calculation.---} We consider the Navier--Stokes equation for an incompressible fluid in $d$ spatial dimensions,
\begin{equation}
    \partial_t \mathbf{v}
    +(\mathbf{v}\cdot\nabla)\mathbf{v}
    = -\nabla P + \nu_0 \nabla^2 \mathbf{v} + \mathbf{f}, \qquad \nabla\cdot \mathbf v=0,
\label{eq:ns}
\end{equation}
driven by a zero-mean transverse Gaussian random force with the pair correlator
\begin{equation}
\begin{aligned}
&\langle f_i(t,\mathbf{k}) f_j(t',-\mathbf{k})\rangle=\delta(t-t')\,D^{\mathrm f}_{ij}(\mathbf{k}),\\
&D^{\mathrm f}_{ij}(\mathbf{k})=D_0\,P_{ij}(\mathbf{k})\,d_{\mathrm f}(k),
\end{aligned}
\label{eq:noise}
\end{equation}
where $k\equiv|\mathbf{k}|$, $P_{ij}(\mathbf k)=\delta_{ij}-k_i k_j/k^2$ is the transverse projector, and $d_{\mathrm f}(k)$ is defined in Eq.~(\ref{eq:pump}). The corresponding action in the Martin--Siggia--Rose formalism~\cite{Martin1973} is multiplicatively renormalizable~\cite{Adzhemyan1983}. A natural expansion parameter is $g_0=D_0/\nu_0^3$.

The central quantity of interest is the correction-to-scaling exponent $\omega$, defined by
\begin{equation}
\omega(\varepsilon) = \left. \frac{\partial \beta(g,\varepsilon)}{\partial g}\right|_{g=g^*}, \quad \beta(g^*,\varepsilon)=0,
\end{equation}
where $\beta(g,\varepsilon)$ is the RG beta function and $g$ is the renormalized counterpart of the bare coupling $g_0$. Owing to the exact relation
\begin{equation}
\beta(g,\varepsilon)=-g\left[ 2\varepsilon - 3\gamma_\nu(g,\varepsilon) \right],
\end{equation}
the calculation of $\omega$ reduces to evaluating the diagrammatic expansion of the anomalous dimension of viscosity $\gamma_\nu(g,\varepsilon)$. Details of the model and its renormalization are given in the End Matter.

The number of diagrams grows rapidly with perturbative order: at five loops, direct enumeration yields more than 56 million diagrams. Their dynamical character is reflected in the presence of time integrations, which, unlike the momentum integrations, can be carried out straightforwardly using the time-version technique~\cite{Vasiliev2004}. Each contributing dynamical diagram is thereby reduced to a set of static momentum integrals.

A substantial reduction in the resulting number of integrals is achieved in the limit $d\to\infty$. As noted in the three-loop study~\cite{Adzhemyan2008}, numerous relations between different momentum integrals arise in this limit. Even without a complete algorithm for identifying all such relations, simply matching integrals that coincide up to relabeling of variables and diagram symmetries reduces the number of independent five-loop integrals to approximately $5.6\times10^5$.

The large-$d$ limit also simplifies the structure of the integrands themselves. To leading order in $1/d$, angular correlations between independent loop momenta are suppressed. For example,
\begin{equation}
\left\langle
\left(
\frac{\mathbf{k}\cdot\mathbf{q}}{kq}
\right)^{2n}
\right\rangle_{S}
=
\mathcal{O}(d^{-n}),
\label{eq:scalar_null}
\end{equation}
where $\langle \dots \rangle_S$ denotes averaging over the unit sphere in $d$ dimensions. Consequently, terms containing nontrivial scalar products of independent momenta as multiplicative factors vanish at leading order in $1/d$. The same rule applies to diagrams with a larger number of loop momenta and greatly simplifies their algebraic structure. Combined with Feynman-parametric integration, this makes an analytic five-loop calculation feasible.

The fifth coefficient in the $\varepsilon$-expansion of $\omega$ was evaluated analytically using a modified version of the algorithm developed for the earlier four-loop calculation~\cite{Adzhemyan2024}. The main tool is parametric integration in terms of hyperlogarithms, as implemented in the \textit{HyperInt} package~\cite{Panzer2015}. For its application, it is convenient to employ a BPHZ-like renormalization scheme~\cite{Brown2013}, in which UV subdivergences are removed by the subtraction $\mathcal R'$-operation with one-scale counterterms. This allows the RG function to be represented in terms of finite combinations of massless integrals suitable for direct parametric integration.

The key requirement is linear reducibility, namely, the existence of an integration order such that the integral remains expressible in terms of hyperlogarithms at every step~\cite{Panzer2014}. Most five-loop integrals can be evaluated without a special choice of integration order, while for the remaining ones linear reducibility is achieved by adjusting the integration order and other elements of the parametric integration procedure. Details of the subtraction procedure and the application of the hyperlogarithm method are given in the Supplemental Material~\cite{SM}. As a check, the algorithm reproduces all known coefficients of the expansion of $\omega$ through four loops. The five-loop calculation yields the coefficient $340/81$ in Eq.~(\ref{eq:omega_series}).

%\emph{\textbf{Conjectured closed form for exponent $\omega$}}

\begin{figure*}
\includegraphics[width=0.70\textwidth]{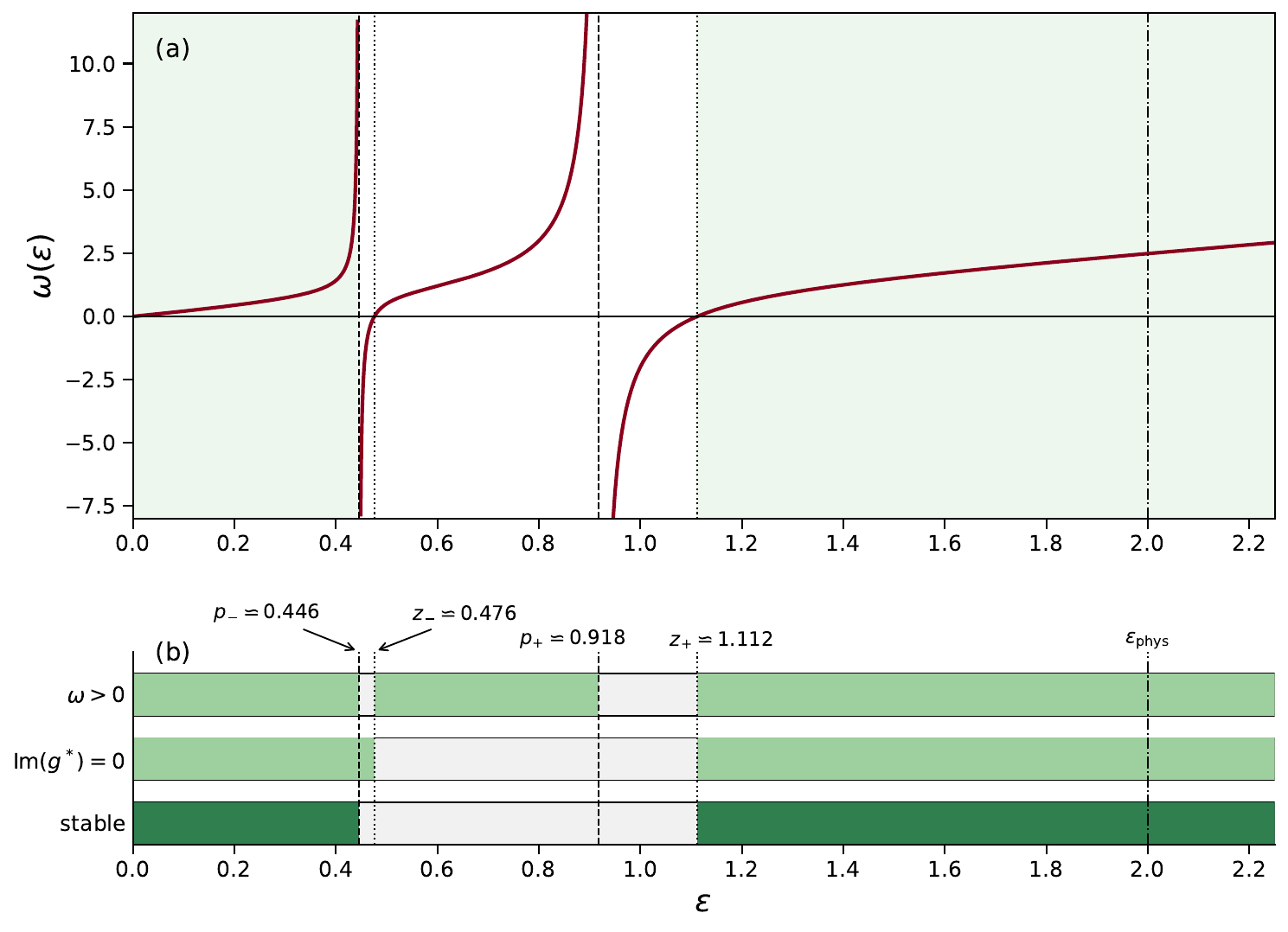}
\caption{
Correction exponent $\omega(\varepsilon)$ and the corresponding stability and reality domains of the RG fixed point.
The upper panel shows the conjectured closed-form expression for $\omega(\varepsilon)$ from Eq.~(\ref{eq:final_omega}).
The lower panel shows the intervals where $\omega>0$, where the fixed point is real, $\operatorname{Im}(g^*)=0$, and where both conditions are simultaneously satisfied.
Vertical lines indicate the corresponding domain boundaries and the physical value $\varepsilon_{\mathrm{phys}}=2$.
}
\label{fig:stability_domains}
\end{figure*}
\textit{Conjectured closed form for exponent $\omega$.---} The rational value of the fifth coefficient supports the conjecture that the expansion of $\omega$ has a simple structure in the large-$d$ limit. To identify a possible closed-form expression for $\omega(\varepsilon)$, we rewrite the expansion as
\begin{equation}
    \omega = 2\varepsilon+\frac{2}{3}\varepsilon^2f(x),
\end{equation}
where
\begin{equation}
    f(x)=1+5x+28x^2+170x^3+\mathcal{O}(x^4),\quad x=\varepsilon/3.
\label{eq:f_series}
\end{equation}

The resulting expansion has integer, rather than merely rational, coefficients. Such integrality is highly unusual in PRG calculations. Given the nontrivial origin of these coefficients, which arise from summing a very large number of diagrams whose individual contributions contain transcendental constants such as Riemann zeta values, it seems unlikely that their integrality is accidental. We therefore conjecture that it reflects an underlying structure of the model and persists to higher orders. This observation motivates us to search the OEIS database~\cite{OEIS} for known sequences reproducing the calculated coefficients. The first three coefficients, $1,5,28$, obtained in the previous four-loop calculation, admit many possible continuations; adding the coefficient $170$ leaves a single matching entry in the OEIS database, A143647~\cite{OEIS:A143647}. It is given by
\begin{equation}
f(x)=\sum_{n=0}^{\infty} a_n x^n,
\quad
a_n=\frac{1}{2}\left[
(5+\sqrt{3})^n+(5-\sqrt{3})^n
\right],
\label{eq:a_n}
\end{equation}
with the generating function
\begin{equation}
f(x)=\frac{1-5x}{1-10x+22x^2}.
\label{eq:f_approx}
\end{equation}
This generating function coincides with the $[1/2]$ Padé approximant to Eq.~(\ref{eq:f_series}). Among the maximal-degree approximants, which reproduce all four known coefficients, the $[1/2]$ approximant is uniquely distinguished by generating an integer sequence that satisfies the simple second-order linear recurrence
\begin{equation}
a_n=10a_{n-1}-22a_{n-2}, \quad n \geq 2.
\end{equation}
A detailed comparison is provided in the Supplemental Material~\cite{SM}.

The selected expression leads to the following conjectured closed form for the exponent $\omega$:
\begin{equation}
\begin{aligned}
    &\omega(\varepsilon)=2\varepsilon + \frac{6\varepsilon^2-10\varepsilon^3}{9-30\varepsilon+22\varepsilon^2} =\frac{17}{11} \varepsilon \frac{(\varepsilon-z_-)(\varepsilon-z_+)}{(\varepsilon-p_-)(\varepsilon-p_+)}, \\
    & z_{\pm} = \frac{3}{34}(9\pm\sqrt{13}), \quad p_{\pm} = \frac{3}{22}(5\pm\sqrt{3}).
    \label{eq:final_omega}
\end{aligned}
\end{equation}

Although it is not a rigorously established result, it continues the integer pattern of the expansion coefficients to higher orders and, as shown below, is consistent with a physical scenario suggested by independent FRG studies. Assuming the validity of Eq.~(\ref{eq:final_omega}), it may be regarded as a conjectured continuation of the perturbative small-$\varepsilon$ result to larger $\varepsilon$ within the RG approach at leading order in $1/d$.

Within this conjectured continuation, we can examine the possible stability and reality domains of the fixed point. For a fixed point to govern the scaling behavior of the system, it must be real and the corresponding exponent $\omega$ must be positive. The condition $\omega>0$ is satisfied in the intervals shown in Fig.~\ref{fig:stability_domains},
\begin{equation}
\begin{aligned}
&0<\varepsilon<p_-\simeq 0.446, \quad  \varepsilon>z_+\simeq 1.112, \\
&z_-\simeq 0.476<\varepsilon<p_+\simeq 0.918
\end{aligned}
\label{eq:epsilon_intervals1}
\end{equation}

To determine whether the fixed point is real, we reconstructed its closed-form expression $g^*(\varepsilon)$ from Eq.~(\ref{eq:final_omega}). This reconstruction is conveniently performed in the minimal subtraction scheme, where the anomalous dimension $\gamma_\nu$ has no explicit dependence on $\varepsilon$; the use of the same $\omega(\varepsilon)$ is justified by the renormalization-scheme invariance of critical exponents. Details of the reconstruction are given in the Supplemental Material~\cite{SM}. The reconstructed fixed-point solution becomes complex in the interval
\begin{equation}
z_-\simeq 0.476
<\varepsilon<
z_+\simeq 1.112.
\end{equation}
Thus, along the continuation from the vicinity of $\varepsilon=0$ to the physical value $\varepsilon_{\mathrm{phys}}=2$, corresponding to physical IR forcing, there are intervals in which the fixed point either ceases to be real or loses stability.

\textit{Discussion.---} The five-loop coefficient preserves the simple rational structure found at lower orders, with the resulting coefficients being not merely rational but integer. It makes it possible to examine a potential continuation of the perturbative result from the vicinity of $\varepsilon=0$ to the physical value $\varepsilon_{\mathrm{phys}}=2$. All nonpolynomial Padé approximants considered, including those of lower total degree, contain a singularity within the interval $0<\varepsilon<2$. Although the position and nature of this singularity depend on how the finite series is continued, its presence in all the approximants considered points to a possible obstruction to the smooth continuation of the real stable branch of the fixed point from small $\varepsilon$ to $\varepsilon_{\mathrm{phys}}$.

This conclusion is obtained in the limit $d\to\infty$, and whether the same mechanism persists at finite $d$ remains an open question. For $d=3$, only the first two orders of the $\varepsilon$-expansion are currently known~\cite{Adzhemyan2003,Adzhemyan2003_2,Adzhemyan2005}, which is insufficient to reveal a structure analogous to that found in the large-$d$ limit. Testing this possibility at finite $d$ would require calculations to substantially higher orders in the full theory, posing a considerably more demanding technical problem.

The $[1/2]$ approximant, singled out by its simple integer and recursive structure, admits a qualitatively distinct physical interpretation. The expression in Eq.~(\ref{eq:final_omega}) has two poles in the interval $\varepsilon \in (0,2)$, giving rise to a second stability region with a real fixed point,
\begin{equation}
\varepsilon>
z_+
\simeq 1.112.
\end{equation}
This region includes the physical value $\varepsilon_{\mathrm{phys}}=2$ and allows the solution to be interpreted as a second real stable branch, separated from the perturbative branch by a region of complex values. Our result does not rigorously establish the existence of a second physically relevant fixed point, but rather points to a possible scenario in which the physical regime is not continuously connected to the perturbative one accessible near $\varepsilon=0$.

This interpretation is consistent with the broader picture emerging from FRG approaches to fully developed turbulence, in which the physical regime driven by forcing at the integral scale is governed by a nonperturbative fixed point that is not directly connected to the perturbative fixed point associated with power-law UV forcing~\cite{FRG_MejaMonasterio2012,FRG_Tarpin2018,FRG_Canet2016,FRG_Canet2022}. Independent evidence for a possible change of regime comes from numerical simulations of turbulence with power-law forcing, where varying $\varepsilon$ was found to induce a transition to an anomalous-scaling regime~\cite{DNS_Sain1998,DNS_Biferale2004,DNS_Biferale2004_2}. Further support for this picture comes from shell models of turbulence, in which distinct fixed points associated with UV and IR forcing have been identified~\cite{Shell_Fontaine2023}.

Although changes in scaling behavior are common in nonequilibrium stochastic systems, partially overlapping intervals in which the same fixed point is unstable and complex-valued are highly unusual. In most cases, the loss of stability of one fixed point is accompanied by the appearance of another stable fixed point that controls the scaling regime. Alternatively, there may exist a finite parameter interval with no stable fixed point and, consequently, no perturbatively accessible scaling regime. A scenario in which the same fixed point becomes unstable over an intermediate range of $\varepsilon$ and subsequently regains stability has been reported only rarely~\cite{Menkyna2020,Antonov2021}. However, these examples of reentrant stability are based on calculations carried out at two-loop order or lower, leaving open the possibility that this behavior is an artifact of low-order perturbation theory. By contrast, if the conjectured exact large-$d$ expression is correct, the overlapping regions of instability and nonreality found here cannot be attributed to a low-order truncation of the RG functions.

Thus, the most robust implication of the five-loop result is the indication of a possible obstruction to smoothly continuing the original perturbative branch to $\varepsilon_{\mathrm{phys}}=2$. The distinguished $[1/2]$ approximant further allows for a second real stable branch near the physical value, although this scenario requires independent verification. The most direct test would be the calculation of the next coefficients in the $\varepsilon$-expansion, which would allow the conjectured closed-form expression to be checked. If the second branch does exist, it remains an open question whether it can be accessed within the PRG framework or whether its emergence is intrinsically tied to the nonperturbative region of the RG flow. 

The existence of a second real domain would also open the possibility of formulating the theory directly at $\varepsilon=2$, with $1/d$ as the only expansion parameter. This would be reminiscent of the $1/N$ expansion~\cite{Vasiliev1982} in the $O(N)$-symmetric $\phi^4$ model~\cite{Kompaniets2017,Schnetz2023}, where the spatial dimension $d$ can be kept arbitrary and the $N\to\infty$ limit is associated with the exactly solvable spherical model~\cite{Stanley1968}. In this sense, establishing a well-defined $d\to\infty$ limit of the turbulence problem at $\varepsilon=2$ could provide an analogous starting point for a systematic $1/d$ expansion. Constructing such an expansion without relying on the $\varepsilon$ parameter would, however, require a reformulation of the theory, which remains an open problem. A possible clue is provided by the unexpectedly large number of linear relations among the integrals encountered in our calculation, which may point to an underlying hidden symmetry or structure of the model.

\section*{Acknowledgments}
We thank Yu. Kirienko for developing several components of an earlier version of the code that are incorporated into the software used in this work. ChatGPT (OpenAI) was used to assist with code implementation. We are also grateful to our colleagues at SPb State University (St. Petersburg, Russia), BLTP JINR (Dubna, Russia),  National Defence University (Helsinki, Finland) and Šafárik University (Košice, Slovakia) for fruitful discussions and comments. The computational algorithms, methodology, and scientific analysis were developed by the authors. D.A.D., D.A.E., and M.V.K.  gratefully acknowledge the support of Foundation for the Advancement of Theoretical Physics ``BASIS" through Grant 25-1-2-48-1.

\section*{Data Availability}
The source code used to generate and evaluate the diagrams is publicly available from~\cite{Code}.

\nocite{Ivanova2023,Cheng1987,Adzhemyan2026,Adzhemyan1998}

\bibliography{references}

\section*{End Matter}
Here we briefly review the renormalization of the stochastic turbulence model. A detailed discussion can be found in~\cite{Adzhemyan1999}.

The stochastic Navier--Stokes equation, Eq.~(\ref{eq:ns}), is
represented in the Martin--Siggia--Rose formalism by the
field-theoretic action
\begin{equation}
S_0=
\frac{1}{2}v_i' D_{ij}^{\mathrm f}v_j'
+
v_i'
\left[
-\partial_t v_i
-(v_j\partial_j)v_i
+\nu_0\nabla^2v_i
\right].
\label{eq:app:action}
\end{equation}
Here $v_i$ and $v_i'$ are the velocity and response fields,
respectively. The noise correlator $D_{ij}^{\mathrm f}$, introduced in Eq.~(\ref{eq:noise}), may in general contain an IR regulator depending on the inverse integral scale of turbulence $m=L^{-1}$:
\begin{equation}
D_{ij}^{\mathrm f}(\mathbf{k}) = D_0 P_{ij}(\mathbf{k}) k^{4-d-2\varepsilon} F\left(\frac{m}{k}\right).
\label{eq:app:noise_cor}
\end{equation}
In the action, all integrations over space and time, as well as summations over repeated indices, are implied. The pressure term in Eq.~(\ref{eq:app:action}) is omitted because the response field $v'_i$ is transverse.

The Feynman diagrams of the perturbation theory generated by the action \eqref{eq:app:action} contain ultraviolet (UV) divergences in the limit $\varepsilon\to0^+$. For $d>2$, such divergences occur only in the one-particle-irreducible correlation function $\langle v_i v_j'\rangle_{\mathrm{1PI}}$. Their cancellation requires a counterterm of the form $v'\nabla^2 v$. The renormalized action is therefore given by
\begin{equation}
S=\frac{1}{2}v_i'D_{ij}^{\mathrm f}v_j'+v_i'\left[-\partial_t v_i-(v_j\partial_j)v_i+\nu Z_\nu\nabla^2v_i\right].
\label{eq:app:action_r}
\end{equation}
It is obtained from \eqref{eq:app:action} by the multiplicative renormalization of the parameters:
\begin{equation}
\begin{aligned}
\nu_0 &= \nu Z_\nu,
&
g_0 &= g\mu^{2\varepsilon}Z_g,
\\
Z_g &= Z_\nu^{-3},
&
D_0 &= g_0\nu_0^3
     = g\mu^{2\varepsilon}\nu^3.
\end{aligned}
\label{eq:app:ren_params}
\end{equation}
Here $\mu$ is the renormalization scale and $g$ is the dimensionless renormalized coupling constant. No field renormalization is required.

In the $(\mathbf{k},t)$ representation, the free propagators of the theory are
\begin{equation}
\begin{aligned}
\langle v_i(t_1)v_j(t_2)\rangle(\mathbf{k})
&=
\frac{g\mu^{2\varepsilon}\nu^2}{2}\,
k^{2-d-2\varepsilon}
e^{-\nu k^2 |t_1-t_2|}
P_{ij}(\mathbf k),\\
\langle v_i(t_1)v'_j(t_2)\rangle(\mathbf{k})
&=
\theta(t_1-t_2)
e^{-\nu k^2(t_1-t_2)}
P_{ij}(\mathbf k),\\
\langle v'_i(t_1)v'_j(t_2)\rangle(\mathbf{k})
&=0 .
\end{aligned}
\label{eq:app:free_propagators}
\end{equation}
In the massless momentum-subtraction scheme used here, we set $m=0$ and assume the normalization $F(0)=1$; the IR regulator is therefore omitted from the propagators. The interaction term in Eq.~(\ref{eq:app:action_r}) generates a cubic vertex, $-v'(v\partial)v = v'_i V_{ijs} v_j v_s/2$, with the vertex factor
\begin{equation}
V_{ijs}= i(k_j \delta_{is} + k_s \delta_{ij}).
    \label{eq:app:vertex}
\end{equation}

Let $\Gamma_{ij}(\mathbf{k},\Omega;\mu)$ denote the unrenormalized one-particle-irreducible (1PI) function $\langle v_i v_j'\rangle_{\mathrm{1PI}}$ calculated with the action \eqref{eq:app:action_r} at $Z_\nu=1$, where $\Omega$ is the frequency conjugate to time. This function is proportional to the transverse projector:
\begin{equation}
\begin{aligned}
\Gamma_{ij}(\mathbf{k},\Omega;\mu)
&=P_{ij}(\mathbf{k})\Gamma(k,\Omega;\mu),\\
\Gamma(k,\Omega;\mu)
&=\frac{\Gamma_{ii}(\mathbf{k},\Omega;\mu)}{d-1}.
\end{aligned}
\label{eq:app:gamma_def}
\end{equation}
We introduce the normalized function
\begin{equation}
\overline{\Gamma}(k,\Omega;\mu) \equiv \frac{\Gamma(k,\Omega;\mu)}{-\nu k^2},
\label{gch}
\end{equation}
which is equal to unity in the tree approximation for $\Omega=0$. Its perturbative expansion is written as
\begin{equation}
\overline{\Gamma}(k,\Omega;\mu) = 1 + \sum_{n\ge 1} {g}^n \mu^{2n\varepsilon} \sum_{\alpha} \chi_n^{(\alpha)}(k,\Omega).
\label{gbar}
\end{equation}
Here $\alpha$ labels the $n$-loop diagrams contributing to
$\overline{\Gamma}$.

The renormalized function $\overline{\Gamma}^{\mathrm R}$, which is free of UV divergences, obeys the RG equation
\begin{multline}
\left[
\mu\partial_\mu
+\beta(g,\varepsilon)\partial_g
-\gamma_\nu(g,\varepsilon)\nu\partial_\nu
\right]
\overline{\Gamma}^{\mathrm R}(k,\Omega;\mu)=
\\
\gamma_\nu(g,\varepsilon)
\overline{\Gamma}^{\mathrm R}(k,\Omega;\mu),
\label{eq:app:rgeq}
\end{multline}
where
\begin{equation}
\gamma_\nu(g,\varepsilon)
=\frac{-2\varepsilon g\partial_g \ln Z_\nu}{1+g\partial_g \ln Z_g}, \quad \beta(g,\varepsilon)=-g\left[2\varepsilon-3\gamma_\nu(g,\varepsilon)\right].
\label{eq:app:beta_gamma_rel}
\end{equation}
The expression for $\beta$ follows from the
relation $Z_g=Z_\nu^{-3}$ in Eq.~(\ref{eq:app:ren_params}). The RG fixed point is determined by the condition $\beta(g^*,\varepsilon)=0$. Its stability is governed by the correction exponent $\omega$,
\begin{equation}
\label{eq:app:omega_def}
\omega(\varepsilon)=\partial_g \beta(g,\varepsilon)|_{g=g^*}\;.
\end{equation}
The fixed point is infrared attractive for $\omega>0$. The Gaussian fixed point, $g^*(\varepsilon)=0$, is unstable since $\omega=-2\varepsilon<0$. A nontrivial fixed point is determined by the condition
\begin{equation}
\gamma_\nu(g^*,\varepsilon)=\frac{2\varepsilon}{3}\;.
\end{equation}
This relation holds exactly and its
$\varepsilon$-expansion contains no higher-order terms.

The relation in Eq.~(\ref{eq:app:beta_gamma_rel}) makes it possible to obtain the correction exponent $\omega$ from the anomalous dimension $\gamma_\nu$. For the purpose of applying parametric integration, it is advantageous to work with finite integrals and to avoid unnecessary mass scales. We therefore express $\gamma_\nu$ in terms of the renormalized function $\overline{\Gamma}^{\mathrm R}$ at the normalization point $\Omega=0$, $k=\mu$, and set the IR scale to zero, $m=0$. The normalization condition is chosen as
\begin{equation}
\overline{\Gamma}^{\mathrm R}|_{k=\mu,\Omega=0} = 1.
\label{eq:app:renorm_con}
\end{equation}
Introducing the dimensionless momentum $\widetilde{k}=k/\mu$ and
hereafter denoting it again by $k$, the RG equation~(\ref{eq:app:rgeq}) evaluated at the normalization point \eqref{eq:app:renorm_con} yields
\begin{equation}
\gamma_\nu(g,\varepsilon)= -(k\partial_k \overline\Gamma^{\mathrm R})|_{k=1,\Omega=0}\;.
\label{gnu3}
\end{equation} 
The finite external momentum at the normalization point provides the required IR regularization.

Subdivergences in the diagrams $\chi^{(\alpha)}$ from Eq.~(\ref{gbar}) are subtracted recursively by Bogoliubov's $\mathcal R'$-operation
\begin{equation}
   \chi^{\mathrm R}=(1-\mathcal K)\mathcal R' \chi, 
\end{equation}
where $\mathcal R'$ removes all UV subdivergences and $(1-\mathcal K)$ subtracts the remaining superficial divergence. For the diagrams considered here, all UV-divergent subgraphs are two-point response subgraphs, whose divergent parts are proportional to their external momentum squared. No overlapping UV-divergent subgraphs occur. The $\mathcal R'$ operation can therefore be written as
\begin{equation}
    \mathcal R' \chi = \prod_s (1-\mathcal K_s) \chi,
\end{equation}
where the index $s$ runs over all divergent subgraphs of the diagram. The operator $\mathcal K_s$ acts on the normalized amplitude of the corresponding subgraph and sets its own external momentum magnitude and frequency to $k_s=1$ and $\Omega_s=0$. For nested subgraphs, the subtraction operators are applied successively from the innermost subgraph to the outermost one.

In the renormalization scheme used here, the subtraction operator is defined by
\begin{equation}
\mathcal K f(k,\Omega)
\equiv
\left.f(k,\Omega)\right|_{k=1,\Omega=0}.
\end{equation}
The $\mathcal K$ operation acts only on superficially divergent diagrams and does not affect the tree-level unity in the perturbative expansion of $\overline{\Gamma}$. Consequently, the renormalized function can be written as
\begin{equation}
\begin{aligned}
\overline{\Gamma}^{\mathrm R}(k,\Omega)
&=(1-\mathcal K)\mathcal R'
\overline{\Gamma}(k,\Omega)
\\
&=\mathcal R'\overline{\Gamma}(k,\Omega)
-\left.
\mathcal R'\overline{\Gamma}(k,\Omega)
\right|_{k=1,\Omega=0}.
\end{aligned}
\label{grsh}
\end{equation}

Finally, the anomalous dimension $\gamma_\nu$ can be written in terms of finite massless integrals as
 \begin{equation}
\gamma_\nu(g,\varepsilon)=  -\sum_{n\ge 1} g^n  \sum_{\alpha} \left. k \partial_k \mathcal R'\chi_n^{(\alpha)}(k,\Omega)\right|_{k=1,\Omega=0}.
\label{eq:app:gamma_exp}
\end{equation} 
Evaluating Eq.~(\ref{eq:app:gamma_exp}) through five loops and combining the result with Eqs.~(\ref{eq:app:beta_gamma_rel}) and (\ref{eq:app:omega_def}), we obtain the five-loop expansion of the exponent $\omega$.

\end{document}

\title{Supplemental Material for\\
``Evidence for an Obstruction to Perturbative Renormalization Group in Stochastic Turbulence''}

\author{L. Ts. Adzhemyan}
\affiliation{Saint Petersburg State University, 7/9 Universitetskaya nab., St. Petersburg, 199034, Russian Federation}
\affiliation{BLTP, Joint Institute for Nuclear Research, 
 Dubna, Moscow Region, 141980,  Russian Federation}

\author{D. A. Davletbaeva}
\affiliation{Saint Petersburg State University, 7/9 Universitetskaya nab., St. Petersburg, 199034, Russian Federation}
\affiliation{Petersburg Nuclear Physics Institute named by B. P. Konstantinov of NRC ``Kurchatov Institute'', mkr. Orlova roshcha 1, Gatchina, Leningradskaya oblast, 188300, Russian Federation}

\author{D. A. Evdokimov}
\affiliation{Saint Petersburg State University, 7/9 Universitetskaya nab., St. Petersburg, 199034, Russian Federation}
\affiliation{BLTP, Joint Institute for Nuclear Research, 
 Dubna, Moscow Region, 141980,  Russian Federation}

\author{M. V. Kompaniets}
\email{m.kompaniets@spbu.ru}
\affiliation{Saint Petersburg State University, 7/9 Universitetskaya nab., St. Petersburg, 199034, Russian Federation}
\affiliation{BLTP, Joint Institute for Nuclear Research, 
 Dubna, Moscow Region, 141980,  Russian Federation}

\maketitle

\section*{Simplifications in the large-$d$ limit}
It is well known that the large-$d$ limit leads to a substantial reduction in the number of diagrams, since many diagrams vanish in this limit~\cite{Adzhemyan2008}. There are two reasons for this reduction. First, angular correlations are suppressed as $d\to\infty$, as shown in Eq.~(7) of the main text; hence, any diagram containing a scalar product of distinct loop momenta as an overall factor vanishes. Second, because of the factor $(d-1)^{-1}$ in the definition of the scalar part of the 1PI function, Eq.~(23), a nonzero result is possible only for diagrams whose contribution to the trace $\Gamma_{ii}(\mathbf{k},\Omega;\mu)$ is of order $\mathcal{O}(d)$. These vanishing conditions can be conveniently formulated as selection rules already at the stage of diagram generation, which substantially simplifies the subsequent calculation. A detailed description of the selection algorithm is given in~\cite{Adzhemyan2008}.

After the nonvanishing diagrams have been selected, each of them generates a set of time versions, that is, momentum integrals obtained after explicit integration over the time variables~\cite{Ivanova2023}. In the model considered here, this integration can be carried out directly because of the simple time dependence of the free propagators, Eq.~(21) in the main text. The reduction of dynamical diagrams to a set of static momentum integrals is a standard technique in multiloop calculations in critical dynamics and nonequilibrium models; for an example of its implementation in the present model, see~\cite{Adzhemyan2024}. The number of time versions grows rapidly with the loop order. In particular, at five loops, each nonvanishing diagram generates on average about 193 time versions. These time versions are then evaluated in the subsequent calculation.

In addition to the vanishing of a large number of diagrams, numerous linear relations are known to arise among the time versions of the remaining diagrams~\cite{Adzhemyan2008}. At present, however, no generalization of the IBP algorithm that would allow such relations to be identified systematically is available for the integrals associated with time versions. Their origin and a general strategy for finding them therefore remain unclear, and the problem becomes particularly difficult at five loops. Moreover, at three loops it was shown that such relations are more readily detected in a massive renormalization scheme, whereas the present calculation employs a massless scheme. We therefore restricted ourselves to the most direct step, namely, the identification of equivalent integrals. Even this simple reduction decreases the number of time versions requiring explicit evaluation by more than an order of magnitude, from $8.49\times10^6$ to $ 5.57\times10^5$ at five loops.
\begin{table}[!ht]
\caption{\label{tab:supp:reduction}
Number of diagrams and their time versions at different loop orders $\ell$. Here, $N_{\mathrm{diag}}$ is the number of diagrams before selection, $N_{\mathrm{diag}}^{\mathrm{nz}}$ is the number of diagrams that remain nonvanishing in the limit $d\to\infty$, $N_{\mathrm{tv}}^{\mathrm{nz}}$ is the number of corresponding time versions, and $N_{\mathrm{tv}}^{\mathrm{uniq}}$ is the number of distinct time versions after the additional identification of identical integrals.
}
\begin{ruledtabular}
\begin{tabular}{ccccc}
$\ell$ &
$N_{\mathrm{diag}}$ &
$N_{\mathrm{diag}}^{\mathrm{nz}}$ &
$N_{\mathrm{tv}}^{\mathrm{nz}}$ &
$N_{\mathrm{tv}}^{\mathrm{uniq}}$ \\
\hline
2 & 60 & 6 & 12 & 7 \\
3 & 4080 & 83 & 520 & 168 \\
4 & 417\,872 & 1692 & 49840 & 7523 \\
5 & 56\,392\,128 & 44062 & 8\,491\,680 & 557\,346 \\
\end{tabular}
\end{ruledtabular}
\end{table}

A further simplification in the large-$d$ limit, which substantially facilitates calculations at higher orders, arises from the structure of the integrands themselves. As shown in Ref.~\cite{Adzhemyan2008}, the number of propagators $\langle v_i v_j\rangle_0$ in any diagram coincides with the number of its loops. The independent loop momenta $\{\mathbf q_r\}_{r=1}^\ell$ can therefore be chosen such that each of them flows through one of these lines. Upon switching to spherical coordinates, the angular dependence drops out of the integrand in the limit $d\to\infty$. Moreover, the factor $q^{2-d-2\varepsilon}$ from the propagator in Eq.~(21) of the main text is compensated by the factor $q^{d-1}$ from the Jacobian:
\begin{equation}
\int \frac{\mathrm{d}^d\mathbf{q}}{(2\pi)^d}\, q^{2-d-2\varepsilon}\ldots
=
\frac{S_d}{(2\pi)^d}\int_0^\infty \mathrm{d}q\, q^{1-2\varepsilon}\ldots.
\label{eq:sup_larged:1}
\end{equation}
Here, $S_d=2\pi^{d/2}/\Gamma(d/2)$ is the surface area of the unit sphere in $d$ dimensions. Thus, after factoring out the angular measure, the resulting integrals contain no explicit dependence on the original spatial dimension $d$.

\section*{Feynman parametrization}
Before introducing the Feynman parametric representation, a set of subtractions for UV subdivergences is constructed for each time version using the recursive $\mathcal{R}'$-operation. In the present work, this part of the algorithm is unchanged from Ref.~\cite{Adzhemyan2024}, where it is described in detail, and we therefore do not repeat its construction here. The essential point is that, after the application of the $\mathcal{R}'$-operation, the original time version is replaced by a set of terms comprising the original time version itself and the corresponding subtraction terms. A separate Feynman parametrization is then constructed for each term. Since all terms share the same general rational structure, they are treated using a common procedure for parametrization and subsequent integration. We briefly describe this procedure below; detailed examples of its application can be found in~\cite{Adzhemyan2024}.

In the limit $d\to\infty$, the integrals under consideration take the form
\begin{equation}
I^{(\ell)} (k)
=
h^\ell
\prod_{i=1}^{\ell}
\int_0^\infty \mathrm{d}q_i\, q_i^{1-2\varepsilon}
\,
\frac{
\prod_{\alpha=1}^{\ell-1} A_\alpha(\{q\},k)
}{
\prod_{\beta=1}^{2\ell-1} B_\beta(\{q\},k)
}.
\label{eq:sup_int:1}
\end{equation}
Here $\ell$ is the number of loops, and, for convenience, the coupling is redefined as
\begin{equation}
h\equiv \frac{gS_d}{(2\pi)^d}.
\label{eq:sup_int:2}
\end{equation}
The factors $A_\alpha$ arise from contractions of the vertex factors, whereas the factors $B_\beta$ result from integration over the time variables. Both are polynomials in the squared momenta.

Since the integrands become independent of the angular variables in the limit $d\to\infty$, the radial momentum integrals can be formally interpreted as angularly averaged integrals in an effective dimension $D$, defined as
\begin{equation}
D=2-2\varepsilon .
\label{eq:sup_int:3}
\end{equation}
In this case, the one-dimensional integration over the momentum magnitude can be written in the equivalent $D$-dimensional form:
\begin{equation}
\int_0^\infty \mathrm{d}q\, q^{1-2\varepsilon} F(q^2)
=
\int \frac{\mathrm{d}^D\mathbf{q}}{S_D} F(\mathbf{q}^2).
\label{eq:sup_int:4}
\end{equation}
Accordingly, Eq.~(\ref{eq:sup_int:1}) can be rewritten as
\begin{equation}
I^{(\ell)} (k)
=
h^\ell
\prod_{i=1}^{\ell}
\left(
\int \frac{\mathrm{d}^D\mathbf{q}_i}{S_D}
\right)
\frac{
\prod_{\alpha=1}^{\ell-1} A_\alpha(\{\mathbf{q}\},k)
}{
\prod_{\beta=1}^{2\ell-1} B_\beta(\{\mathbf{q}\},k)
}.
\label{eq:sup_int:5}
\end{equation}

To introduce the Feynman parametric representation, we use the identity
\begin{equation}
\frac{
\prod_{\alpha=1}^{\ell-1} A_\alpha
}{
\prod_{\beta=1}^{2\ell-1} B_\beta
}
=
\Gamma(\ell)
\left.
\prod_{\alpha=1}^{\ell-1}
\left(-\frac{\partial}{\partial u'_\alpha}\right)
\int_0^1
\prod_{\beta=1}^{2\ell-1}\mathrm{d}u_\beta\,
\frac{
\delta\!\left(1-\sum_{\beta=1}^{2\ell-1}u_\beta\right)
}{
H(\{\mathbf{q}\},k;u,u')^\ell
}
\right|_{u'=0},
\label{eq:sup_int:6}
\end{equation}
where $u=(u_1,\ldots,u_{2\ell-1})$ are the Feynman parameters, and $u'=(u'_1,\ldots,u'_{\ell-1})$ are auxiliary parameters associated with the numerator factors. The quantity $H$ is defined as
\begin{equation}
H(\{\mathbf{q}\},k;u,u')
=
\sum_{\beta=1}^{2\ell-1} u_\beta B_\beta(\{\mathbf{q}\},k)
+
\sum_{\alpha=1}^{\ell-1} u'_\alpha A_\alpha(\{\mathbf{q}\},k) .
\label{eq:sup_int:7}
\end{equation}
It can be written as a quadratic form in the momenta:
\begin{equation}
H(\{\mathbf{q}\},k;u,u')
=
\sum_{i,j=1}^{\ell}
v_{ij}(u,u')\,
\mathbf{q}_i\!\cdot\!\mathbf{q}_j
+
c(k;u,u') .
\label{eq:sup_int:8}
\end{equation}
The first Symanzik polynomial is given by the determinant of this quadratic form:
\begin{equation}
\mathcal{U}(u,u')=\det v(u,u') .
\label{eq:sup_int:9}
\end{equation}
At leading order in $1/d$, scalar products between distinct independent loop momenta can be neglected in the quadratic forms entering the denominators. With the momentum routing introduced above, the matrix $v$ is therefore diagonal.

Performing the momentum integrations in Eq.~(\ref{eq:sup_int:5}) at $D=2-2\varepsilon$, we obtain:
\begin{equation}
\prod_{i=1}^{\ell}
\int \frac{\mathrm{d}^D\mathbf{q}_i}{S_D}
\frac{1}{H^\ell}
=
\frac{
\Gamma^\ell(1-\varepsilon)\,
\Gamma(\ell\varepsilon)
}{
2^\ell\Gamma(\ell)
}
\frac{1}{
c(k;u,u')^{\ell\varepsilon}\,
\mathcal{U}(u,u')^{1-\varepsilon}
}.
\label{eq:sup_int:10}
\end{equation}
Combining Eq.~(\ref{eq:sup_int:6}) with Eq.~(\ref{eq:sup_int:10}), we obtain the general parametric representation for the integrals under consideration:
\begin{equation}
I^{(\ell)}(k)
=
h^\ell
\frac{
\Gamma^\ell(1-\varepsilon)\,
\Gamma(\ell\varepsilon)
}{
2^\ell
}
\left.
\prod_{\alpha=1}^{\ell-1}
\left(-\frac{\partial}{\partial u'_\alpha}\right)
\int_0^1
\prod_{\beta=1}^{2\ell-1}\mathrm{d}u_\beta\,
\frac{
\delta\!\left(1-\sum_{\gamma=1}^{2\ell-1}u_\gamma\right)
}{
c(k;u,u')^{\ell\varepsilon}\,
\mathcal{U}(u,u')^{1-\varepsilon}
}
\right|_{u'=0}.
\label{eq:sup_int:11}
\end{equation}
According to the Cheng--Wu theorem~\cite{Cheng1987}, the $\delta$-function in the integral can be eliminated by fixing any one of the integration parameters to unity and extending the integration range of all remaining parameters to $(0,\infty)$. An appropriate choice of the fixed parameter is then used to achieve linear reducibility for integrals that are not linearly reducible under the original choice. Representation~(\ref{eq:sup_int:11}), together with the Cheng--Wu transformation described above, serves as the starting point for the subsequent parametric integration.

\section*{Algorithm of parametric integration}
After constructing the subtraction terms that remove subdivergences and converting all integrals to the Feynman parametric representation, we evaluated them using the \textit{HyperInt} package, which implements the algorithm of parametric integration~\cite{Panzer2015}. A potential obstruction is the lack of linear reducibility, i.e., the impossibility of expressing the integrand in terms of hyperlogarithms after each integration step. Unlike in the previous four-loop calculation~\cite{Adzhemyan2024}, integrals that are not linearly reducible arise at five loops, requiring the modification of the computational procedure described below.

Although the complete $\mathcal{R'}$-subtracted combination is finite, its individual terms may contain poles in $\varepsilon$. These terms can be evaluated separately, provided that the same parametrization and integration prescription are used throughout; the poles then cancel in their sum. Therefore, all integrals associated with a given time version must be evaluated consistently, using the same Feynman parameter fixed to unity and the same integration order.

For each time version containing non-linearly-reducible integrals, we tested different integration orders and choices of the parameter fixed to unity. As shown in Ref.~\cite{Adzhemyan2026} for multiloop integrals with a similar structure of dynamical propagators, an appropriate choice of this Feynman parameter can substantially accelerate the calculation, yield hyperlogarithms with a simpler alphabet, and, in some cases, restore linear reducibility. The preferred choice is the parameter associated with the denominator factor containing the largest number of distinct loop momenta. If several such factors are present, the remaining candidates are tested sequentially whenever integration fails. The optimal integration order was determined using the polynomial-reduction procedure implemented in \textit{HyperInt}. Without performing the integrations explicitly, this algorithm analyzes the polynomials generated at intermediate steps and selects an order that minimizes their number and complexity.

The complete algorithm for evaluating a time version to the required order in $\varepsilon$, which directly yields its contribution to the anomalous dimension $\gamma_\nu(g,\varepsilon)$ in Eq.~(36) in the main text, is as follows:
\begin{enumerate}
\item In the original integrand for the given time version, identify the denominator factors of the form $q^2_{i_1} + q^2_{i_2} + \cdots +q^2_{i_s}$ containing the largest number of distinct loop momenta, i.e., for which $s$ is maximal.
\item Construct the set of terms generated by the recursive application of the $\mathcal{R}'$ operation: the original term and the subtraction terms that remove subdivergences.
\item For each term, construct the Feynman-parametric representation, differentiate with respect to the auxiliary parameters $u'_\alpha$ and the external momentum $k$, and then set $u'=0$ and $k=1$.
\item Expand the resulting parametric integrand in $\varepsilon$ to the required order.
\item Choose one of the Feynman parameters associated with the denominator factors identified in step 1 and fix it to unity.
\item For the main term and the chosen unit parameter, determine the optimal integration order using polynomial reduction implemented in \textit{HyperInt}.
\item Integrate all terms of the given time version using the same integration order and unit parameter.
\item If some terms remain non-linearly-reducible and cannot be evaluated in the preceding step, return to step 5 and choose another unit parameter.
\item If some terms still cannot be evaluated after all candidate unit parameters have been tested, return to step 6, but determine the optimal integration order from one of the unevaluated terms rather than from the main term.
\end{enumerate}
This procedure was sufficient to obtain a linearly reducible representation for all five-loop time versions encountered in the calculation. The source code used in the calculation is publicly available online~\cite{Code}.

\section*{Comparison of Padé approximants}
Here we consider Padé approximants constructed for the function
\begin{equation}
    f(x)=1+5x+28x^2+170x^3+\mathcal{O}(x^4),\quad x=\varepsilon/3,
\label{eq:sup:f_series}
\end{equation}
defined in Eq.~(9) of the main text. The corresponding approximations for the exponent $\omega(\varepsilon)$ are then obtained as described in the main text. All nonpolynomial approximations for $\omega(\varepsilon)$ possess a pole in the interval $0<\varepsilon<\varepsilon_{\mathrm{phys}}=2$, including those obtained from Padé approximants of lower total degree, as summarized in Table~\ref{tab:supp:pade_poles}. Remarkably, the location of the first positive pole remains close to $\varepsilon\simeq 1/2$ across all approximants considered.
\begin{table}[!ht]
\centering
\begin{minipage}{0.7\textwidth}
\caption{\label{tab:supp:pade_poles}
Positive poles of the Pad\'e approximations for
$\omega(\varepsilon)$ obtained from the corresponding approximants
to $f(x)$. Pole positions are given to three decimal places.
}
\begin{ruledtabular}
\begin{tabular}{cc}
Pad\'e approximant for $f(x)$ &
Poles of $\omega(\varepsilon)$ \\
\hline
$[0/1]$ & $0.600$ \\
$[1/1]$ & $0.536$ \\
$[0/2]$ & $0.541$ \\
$[2/1]$ & $0.494$ \\
$[1/2]$ & $0.446,\;0.918$ \\
$[0/3]$ & $0.506$ \\
\end{tabular}
\end{ruledtabular}
\end{minipage}
\end{table}

At the maximal total degree, the expansion admits three approximants, $[0/3]$, $[1/2]$, and $[2/1]$. We reproduce the $[1/2]$
approximant from Eq.~(11) of the main text and list all three together
for comparison:
\begin{equation}
\begin{aligned}
&    f_{[1/2]}(x)=\frac{1-5x}{1-10x+22x^2}, \\
&    f_{[0/3]}(x) = -\frac{1}{15x^3+3x^2+5x-1}, \\
&    f_{[2/1]}(x) = \frac{33x^2+15x-14}{85x-14}.
\end{aligned}
\label{eq:sup_pade:1}
\end{equation}
They yield the following expressions for the exponent $\omega$:
\begin{equation}
\begin{aligned}
&     \omega_{[1/2]}(\varepsilon)=2\varepsilon + \frac{6\varepsilon^2-10\varepsilon^3}{9-30\varepsilon+22\varepsilon^2}, \\
&    \omega_{[0/3]}(\varepsilon) = 2\varepsilon+\frac{6\varepsilon^2}{9-15\varepsilon-3\varepsilon^2-5\varepsilon^3}, \\
&    \omega_{[2/1]}(\varepsilon) = 2\varepsilon+\frac{22\varepsilon^4+30\varepsilon^3-84\varepsilon^2}{255\varepsilon-126}.
\end{aligned}
\label{eq:sup_pade:2}
\end{equation}

Like the $[1/2]$ approximant, both expressions have a pole near $\varepsilon\simeq 1/2$. The essential difference is that the $[0/3]$ and $[2/1]$ approximants possess no second pole in the interval $\varepsilon\in[0,2]$. If either were adopted as the correct continuation, no additional region of fixed-point stability would emerge that includes the physical value $\varepsilon_{\mathrm{phys}}=2$.

In addition to possessing a second pole, the $[1/2]$ approximant is distinguished by the simpler structure of its expansion coefficients. In particular, unlike the other two approximants, the $[2/1]$ approximant has noninteger coefficients starting at fourth order.
\begin{equation}
    f_{[2/1]}(x) = 1+5x+28x^2+170x^3+\frac{7225}{7}x^4+\mathcal{O}(x^5).
\label{eq:sup_pade:3}
\end{equation}
Since the integer coefficients in Eq.~(\ref{eq:sup:f_series}) result from summing the contributions of millions of distinct time versions, there is reason to regard their integrality as a nontrivial property of the expansion rather than a mere coincidence.

The $[0/3]$ approximant, which yields integer higher-order coefficients, is governed by a third-order linear recurrence relation
\begin{equation}
    a_n=5 a_{n-1} + 3 a_{n-2} + 15a_{n-3}, \quad n\geq3.
\label{eq:sup_pade:4}
\end{equation}
Compared with the $[1/2]$ approximant, this leads to a considerably more complicated expression for the $n$th coefficient:
\begin{equation}
\begin{aligned}
    a_n=\sum_{j=0}^{2} \frac{\lambda^{n+2}_j}{3\lambda_j^2-10\lambda_j-3},
\end{aligned}
\label{eq:sup_pade:5}
\end{equation}
where $\lambda_j$ are the roots of the characteristic equation
\begin{equation}
    \lambda^3-5\lambda^2-3\lambda-15=0.
\label{eq:sup_pade:6}
\end{equation}

Thus, among the two maximal-degree approximants that generate integer higher-order coefficients, the $[1/2]$ approximant has the simpler recurrence and closed-form structure. It is also the only approximant considered that possesses a second pole in the interval $0<\varepsilon<2$, allowing for an additional stability region containing the physical value $\varepsilon_{\mathrm{phys}}=2$.

\section*{Reconstruction of a closed-form fixed point in the MS scheme}
To determine the range of $\varepsilon$ over which the fixed point $g^*(\varepsilon)$ remains real, we reconstruct its closed-form expression from the conjectured closed form of the exponent $\omega$
\begin{equation}
    \omega(\varepsilon)=2\varepsilon + \frac{6\varepsilon^2-10\varepsilon^3}{9-30\varepsilon+22\varepsilon^2}.
    \label{eq:final_omega}
\end{equation}
For convenience, we use the rescaled coupling $h=gS_d/(2\pi)^d$ introduced in Eq.~(\ref{eq:sup_int:2}). Since $h$ differs from $g$ only by a real positive factor independent of $\varepsilon$, the reality domains of $h^*(\varepsilon)$ and $g^*(\varepsilon)$ coincide. This reconstruction cannot be performed within the massless momentum-subtraction scheme used for the calculations, because $\gamma_{\nu}$ depends explicitly on $\varepsilon$. The most direct approach is therefore to switch to the minimal subtraction (MS) scheme, in which the anomalous dimension $\gamma_{\nu}(h)$ has no explicit dependence on $\varepsilon$. This change of scheme is legitimate because critical exponents are invariant under changes of the renormalization scheme. In what follows, all quantities in the MS scheme are denoted by tildes.

Using the relation between the $\beta$ function and the anomalous dimension $\gamma_\nu$ (Eq.~(6) of the main text), we obtain the following expression for the exponent $\omega$:
\begin{equation}
    \omega(\varepsilon)=\left. \frac{\partial \tilde{\beta}(\tilde{h},\varepsilon)}{\partial \tilde{h}}\right|_{\tilde{h}=\tilde{h}^*} =  -2\varepsilon + 3\tilde{\gamma}_{\nu}(\tilde{h}^*) + 3\tilde{h}^*(\varepsilon)  \left.\frac{\partial \tilde{\gamma}_{\nu}(\tilde{h})}{\partial \tilde{h}} \right|_{\tilde{h}=\tilde{h}^*}.
    \label{eq:sup_MS:1}
\end{equation}
On the other hand, the fixed-point condition $\tilde{\beta}(\tilde{h}^*,\varepsilon)=0$ implies
\begin{equation}
    \tilde{\gamma}_{\nu}(\tilde{h}^*) = \frac{2}{3} \varepsilon.
    \label{eq:sup_MS:2}
\end{equation}
Differentiating this relation with respect to $\varepsilon$ and using the fact that, in the MS scheme, $\tilde{\gamma}_{\nu}$ depends on $\varepsilon$ only through $\tilde{h}^*(\varepsilon)$, we obtain
\begin{equation}
   \left. \frac{\partial\tilde{\gamma}_{\nu}(\tilde{h})}{\partial \tilde{h}}\right|_{\tilde{h}=\tilde{h}^*} = \frac{2}{3} \left( \frac{d\tilde{h}^*(\varepsilon)}{d\varepsilon} \right)^{-1}.
   \label{eq:sup_MS:3}
\end{equation}

Substituting Eqs.~(\ref{eq:sup_MS:2}) and~(\ref{eq:sup_MS:3}) into Eq.~(\ref{eq:sup_MS:1}) yields the differential equation
\begin{equation}
    \frac{d \tilde{h}^*}{\tilde{h}^*} = \frac{2d\varepsilon}{\omega(\varepsilon)}.
    \label{eq:sup_MS:4}
\end{equation}
Integrating this equation and fixing the integration constant from the known leading term of the fixed-point expansion in the MS scheme~\cite{Adzhemyan2008},
\begin{equation}
    \tilde h^*(\varepsilon) = \frac{8}{3}\varepsilon + \mathcal{O}(\varepsilon^2),
\label{eq:sup_MS:5}
\end{equation}
we obtain the desired closed-form expression:
\begin{equation}
\tilde{h}^*(\varepsilon) = \frac{8}{3}\varepsilon \left[ \left(1-\frac{\varepsilon}{z_-}\right) \left(1-\frac{\varepsilon}{z_+}\right) \right]^{\frac{5}{34}} \left[ \frac{1-\varepsilon/z_+}{1-\varepsilon/z_-} \right]^{\frac{11}{34\sqrt{13}}},
\label{eq:epsilon_intervals2}
\end{equation}
where $z_{\pm}$ are defined in Eq.~(13) of the main text. This expression is used to obtain the real-valued domains of the fixed point, depicted in Fig.~1 in the main text.

We note that, although fixed-point coordinates are scheme dependent, the existence of a real fixed point is invariant under admissible changes of renormalization scheme. The corresponding reality domains in $\varepsilon$ should therefore be the same in the MS and momentum-subtraction schemes~\cite{Adzhemyan1998}.

\bibliography{references}